\documentclass[prd,superscriptaddress,floatfix,notitlepage,nofootinbib,reprint]{revtex4-1} 

\usepackage{graphicx,amsmath,amssymb,bm}
\usepackage{psfrag}
\usepackage{feynmp}
\usepackage{hyperref}
\usepackage{slashed}
\usepackage{subcaption}

   \newcommand{\exclude}[1]{}
\newcommand{\be}{\begin{eqnarray}}
\newcommand{\ee}{\end{eqnarray}}
\renewcommand{\d}{\mathrm{d}}

{\catcode`\|=\active
  \gdef\Braket#1{\left<\mathcode`\|"8000\let|\bravert {#1}\right>}}

\newcommand{\ISM}{\textsc{ism}}

\providecommand{\d}{}
\renewcommand{\d}{{\rm d}}
\providecommand{\d}{}
\renewcommand{\d}{{\rm d}}

\newcommand{\beq}{\begin{equation}}
\newcommand{\eeq}{\end{equation}}

\def\ra{\rangle}
\def\la{\langle}

\begin{document}
  
  \title{Primordial Lithium Puzzle and the Axion Quark Nugget  Dark Matter Model}
       \author{Victor V. Flambaum}
       \email{v.flambaum@unsw.edu.au}
        \affiliation{School of Physics, University of New South Wales, Sydney 2052, Australia}
       \author{Ariel  R. Zhitnitsky}
\email{arz@phas.ubc.ca}
 \affiliation{Department of Physics and Astronomy, University of British Columbia, Vancouver, V6T 1Z1, BC, Canada}
     
      \begin{abstract} 

Astrophysics today faces a number of mysteries which defiant their resolutions   in spite  of drastic improvements in instrumental  design, better technique being developed, and 
 gradual improvements in theoretical and computation methods 
over last decades. Primordial Lithium Puzzle is known to stay with us for at least two decades,     and it is very likely  that its final resolution   will  require 
some fundamentally new ideas, novel  frameworks and a non-conventional paradigms. 
 We propose that Primordial Lithium Puzzle  finds its natural resolution 
 within the so-called  Axion Quark Nugget (AQN) dark matter model. This model was invented long ago as a natural explanation of the observed ratio $\Omega_{\rm dark} \sim   \Omega_{\rm visible}$ without any references   to BBN physics. In this new paradigm, in contrast with conventional  WIMP (Weakly Interacting Massive Particle) framework   the dark matter (DM) takes the form  of the macroscopically large quark nuggets without requiring any new fields beyond the standard model physics, except for the axion. The time evolution of these AQNs in primordial soup at $T\sim 20$ KeV suggests a strong suppression of the  
 abundances of
  nuclei   with high charges $Z\geq 3$. This suppression mechanism    represents the resolution   of the primordial lithium abundance within AQN dark matter scenario.

   \end{abstract}
 \maketitle

 \flushbottom
\section{Introduction}
  Prediction of the primordial abundances of light elements during the Big Bang Nucleosyntheis (BBN) is one of the major triumphs  of physics and cosmology. Indeed, the BBN theory has no free parameters to fit the data (if the barion-to-photon ratio $\eta$ is taken from the Cosmic Microwave Background data). In spite of the complexity of many coupled  nuclear reactions during BBN the abundances of helium and deuterium  (which differs by 4 orders of magnitude) are predicted with high accuracy. The only remaining problem is the abundance of $^7$Li which is predicted about 3 times larger than the results of the observations \cite{ParticleData}.

  In this paper we propose a specific mechanism on resolution of the  $^7$Li  puzzle within  the  Axion Quark Nugget (AQN) dark matter model, see next section \ref{sec:AQN} with  a short overview of this model. The abundance of $^7$Li  is build up from the direct $^7$Li  production and by the decay of $^7$Be nucleus to $^7$Li. Both nuclei have relatively high charge $Z=3$ and $Z=4$. In this paper we show that a finite portion of these high charge nuclei might be  captured and subsequently annihilated by the  negatively charged antinuggets which provide strong attraction for the positively charged $^7$Be and $^7$Li nuclei. The dependence on the nuclear charge is exponential, therefore, the abundances of lighter nuclei ($^4$He, $^3$He, $^2$H and $^1$H) are not affected.
  
  We refer to recent review paper  \cite{BBN-review} with detail discussions on possible paths on resolutions to the   Primordial Lithium Problem, which are classified by  ref.\cite{BBN-review} as follows: \\1. Astrophysical Solutions;\\ 2. Nuclear Physics Solutions; \\3. Beyond the Standard Model   Solutions. 
  
  Our proposal does not literally belong to any of these categories  as the crucial element of the proposal is the quark nuggets representing the  dark matter   made of conventional quarks and gluons from the Standard  Model,
  (though in   a different, not conventional hadronic, phase). Still our proposal  can be vaguely  classified by category 3 as there is a new element in the AQN model, the axion,  which is not a part of the Standard Model, yet.

 The paper is organized as follows. In next section \ref{sec:AQN} we overview the AQN model by paying special attention to the astrophysical and cosmological consequences of this specific dark matter  model.    In Section \ref{sec:electrosphere} we   overview some technical details related to the internal structure of the nuggets, which plays an important  role in context of the present work on the AQN induced suppression of the BBN nuclei with $Z\geq 3$. In Section \ref{sec:suppression} we describe the mechanism which we think 
  is capable to   suppress the  primordial lithium abundance during and shortly after the BBN. We conclude in Section \ref{conclusion} with few thoughts on the future developments and possible tests of this proposal.

 \section{Axion Quark Nugget (AQN)  model}
\label{sec:AQN}

The AQN Model in the title of this section stands for the axion quark nugget   model to emphasize on essential role of the axion field and avoid confusion with earlier models, see below. This title includes two very different notions: the ``axion" and the ``quark nugget".  

We start with the term ``axion". We refer to the original papers \cite{axion,KSVZ,DFSZ} on the axion field   and the recent activities related to the axion search experiments \cite{vanBibber:2006rb,Asztalos:2006kz,Sikivie:2008,Raffelt:2006cw,Sikivie:2009fv,Rosenberg:2015kxa,Graham:2015ouw,Marsh:2015xka,Ringwald:2016yge,Irastorza:2018dyq}.  We continue with the term ``quark nugget". 
 The idea that the dark matter may take the form of composite objects of 
standard model quarks in a novel phase goes back to quark nuggets  \cite{Witten:1984rs}, strangelets \cite{Farhi:1984qu}, nuclearities \cite{DeRujula:1984axn},  see also application of this idea  to strange stars \cite{Alcock:1986hz, Kettner:1994zs}, and  review \cite{Madsen:1998uh} with large number of references on the original results. 
In the models \cite{Witten:1984rs,Farhi:1984qu,DeRujula:1984axn,Alcock:1986hz, Kettner:1994zs,Madsen:1998uh}  the presence of strange quark stabilizes the quark matter at sufficiently 
high densities allowing strangelets being formed in the early universe to remain stable 
over cosmological timescales.  There were a number of problems with the original idea\footnote{\label{first-order}In particular, the first order phase transition is a required feature of the system for the strangelet  to be formed during the QCD phase transition.  However it is known by now that the QCD transition is a crossover rather than the first order phase transition as the recent lattice results \cite{Aoki:2006we} unambiguously show. Furthermore, the strangelets 
will likely evaporate on the Hubble time-scale even if they had been formed \cite{Alcock:1985}.} and we refer to the review paper \cite{Madsen:1998uh} for the details. 

The quark nugget model advocated in \cite{Zhitnitsky:2002qa} is conceptually similar, with the 
nuggets being composed of a high density colour superconducting (CS) phase.
An additional stabilization factor in the quark nugget model is provided by the axion domain walls
  which are copiously produced during the QCD transition\footnote{In this case the first order phase transition is not required for the nuggets to be formed as the axion domain wall plays the role of the squeezer. Furthermore, the argument  related to the fast evaporation of the strangelets as mentioned in  footnote \ref{first-order}   is not applicable for the  quark nugget model \cite{Zhitnitsky:2002qa} because the  vacuum ground state energies inside (CS phase) and outside (hadronic phase)  the nuggets are drastically different. Therefore these two systems can coexist only in the presence of the additional external pressure provided by the axion domain wall, in contrast with strangelet models \cite{Witten:1984rs,Madsen:1998uh} which  must  be stable at zero external pressure.}.
 The crucial  novel  additional  element in the proposal \cite{Zhitnitsky:2002qa} (in addition to the presence of the axion domain wall)  is that the nuggets could be 
made of matter as well as {\it antimatter} in this framework.

This novel  key element of the model \cite{Zhitnitsky:2002qa} completely changes entire framework 
because the dark matter density  $\Omega_{\rm dark}$ and the baryonic matter density $ \Omega_{\rm visible}$ now become intimately related
to each other and proportional to each other.
Indeed, the conservation of the baryon charge implies
 \be
  \label{eq:321}
    B_{\text{universe}} = 0 &=& B_{\text{nugget}}
    + B_{\text{visible}}- |{B}|_{\text{antinugget}}\nonumber \\
    |B|_{\text{dark-matter}} &=& B_{\text{nugget}} +
   |{B}|_{\text{antinugget}}  
\ee
where $B_{\text{universe}}=0$ is the total number of baryons
 in the universe,
$|B|_{\text{dark-matter}}$ counts  total number of baryons  
and  total number of antibaryons hidden in the nuggets and
antinuggets that make up the dark matter,  and
$B_{\text{visible}}$ is the total number of residual ``visible''
baryons (regular matter).   The energy per baryon charge   is approximately the same 
for nuggets and the visible matter  as the both types of matter are formed during the same QCD transition, and both are proportional to the same dimensional  parameter  $\sim m_p$ which  implies that 
\be
\label{Omega}
 \Omega_{\rm dark}\sim \Omega_{\rm visible}
\ee
see recent  refs.   \cite{Liang:2016tqc,Ge:2017ttc,Ge:2017idw} for the  details. In other words, the nature of dark matter and the problem of the asymmetry between matter and antimatter in the Universe, normally formulated as the so-called baryogenesis problem,   becomes  two sides of the same coin in this framework.  As it has been  argued in refs.   \cite{Liang:2016tqc,Ge:2017ttc,Ge:2017idw}  the relation   (\ref{Omega})  is  very generic outcome of the AQN framework, and it is  not sensitive to any specific details of the model.  

The AQN proposal  represents an 
alternative to baryogenesis  scenario  when  the 
``baryogenesis'' is replaced by  a charge separation process 
in which the global baryon number of the Universe remains 
zero. In this model the unobserved antibaryons  come to comprise 
the dark matter in the form of dense antinuggets in colour superconducting (CS) phase.  The dense nuggets in CS phase also present in the system
such that the total baryon charge remains zero at all times during the evolution of the Universe. The detail mechanism of the formation of the  nuggets 
and antinuggets has been recently developed in refs.   \cite{Liang:2016tqc,Ge:2017ttc,Ge:2017idw}. We highlight below the basics elements of this proposal, its predictions and the observational consequences including presently available constraints. 
 
\begin{figure}
\centering
\captionsetup{justification=raggedright}
\includegraphics[width=0.8\linewidth]{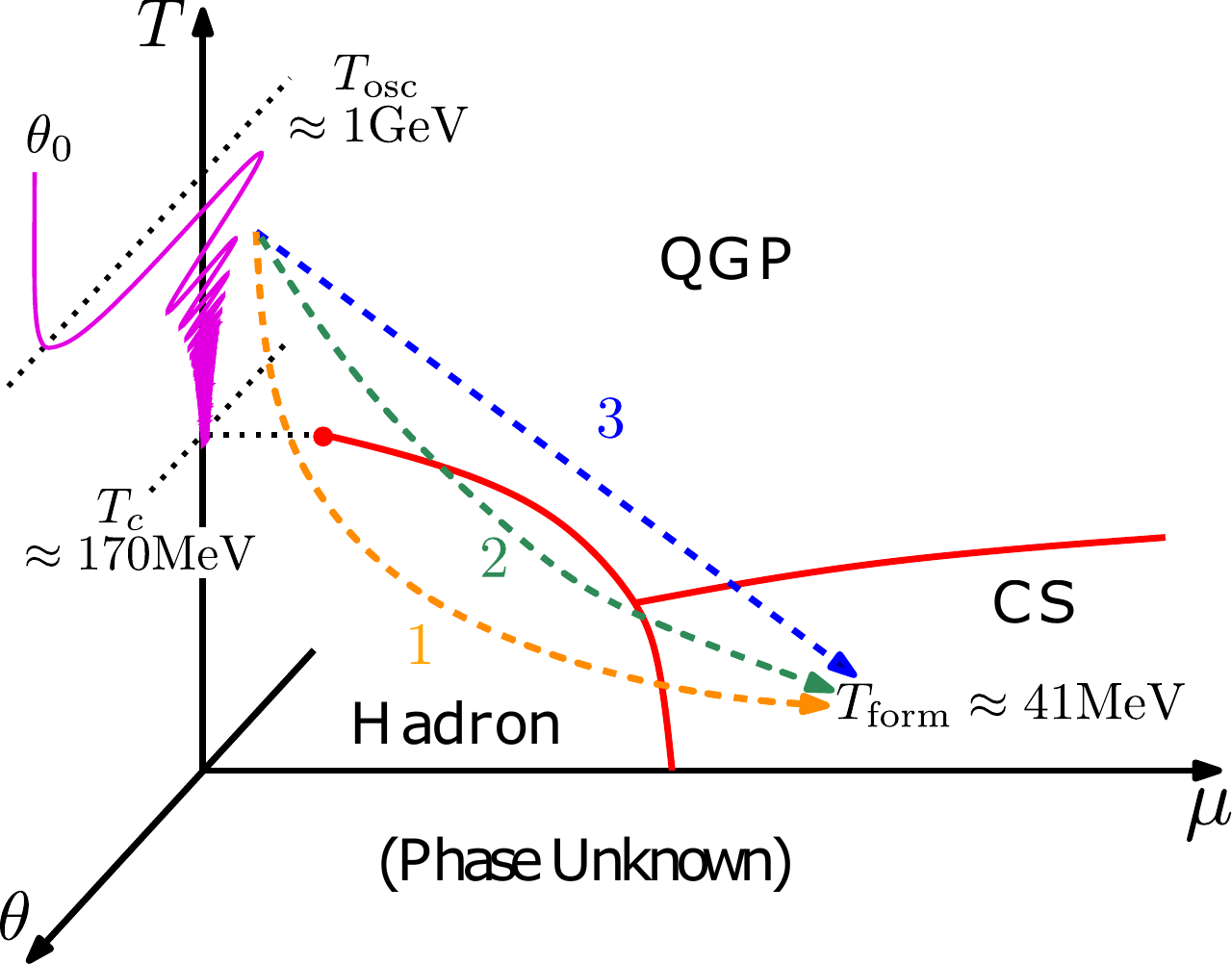}
\caption{This diagram illustrates the interrelation between the axion production due to the misalignment mechanism and the nugget's formation which starts before the axion field $\theta$ relaxes to zero. Adopted from \cite{Ge:2017idw}.}
\label{phase_diagram}
\end{figure}

If the fundamental $\theta$ parameter of QCD were identically zero  
during the formation time, see Fig. \ref{phase_diagram}, than equal numbers of nuggets 
made of matter and antimatter would be formed.  However, the fundamental $\cal CP$ violating processes associated 
with the $\theta$ term in QCD   result in the preferential formation of 
antinuggets over nuggets.   This source of strong $\cal CP$ violation is no longer available at the present epoch as a result of the axion dynamics  when $\theta$ eventually relaxes to zero as a result of the axion dynamics.
    Due to this global  $\cal CP$ violating processes during the early formation stage  the number of nuggets and antinuggets 
      being formed would be different. This difference is always of order of one effect   irrespectively to the parameters of the theory, the axion mass $m_a$ or the initial misalignment angle $\theta_0$, as argued in  \cite{Liang:2016tqc,Ge:2017ttc}.   As a result of this disparity between nuggets and antinuggets   a similar disparity would also emerge between visible quarks and antiquarks according to (\ref{eq:321}).  Precisely this disparity between visible baryons  and antibaryons eventually lead (as a result of the annihilation processes) to the system when exclusively one  species of visible baryons  remain in the system,  in agreement with observations.
     
      One should emphasize that this global  $\cal CP$ violation is correlated on enormous scales of the entire visible Universe because in this framework it is assumed that the inflation occurs after Peccei- Quinn (PQ) phase transition with the  scale $f_a$. Nevertheless, the so-called $N_{\rm DW}=1$ domain walls (which correspond to the interpolation between one and the same  unique vacuum state)
     can be formed at the QCD transition even if the  inflation occurs after PQ scale, i.e. $T_{\rm infl}< f_a$ and, therefore,  entire visible Universe is characterized by unique $\theta$ vacuum state, see  \cite{Liang:2016tqc} with detail discussions and arguments
     supporting this claim. Precisely these  $N_{\rm DW}=1$ domain walls play a key role in formation  of the nuggets.

      The disparity between nuggets and antinuggets  unambiguously implies that  the total number of  visible antibaryons  will be less than the number of baryons in   early universe plasma as  (\ref{eq:321}) states.
 This  is precisely  the reason why the resulting visible and dark matter 
densities must be the same order of magnitude (\ref{Omega})  in this framework  
as they are both proportional to the same fundamental $\Lambda_{\rm QCD} $ scale,  
and they both 
 originated at the same  QCD epoch.
  If these processes 
are not fundamentally related, the two components 
$\Omega_{\rm dark}$ and $\Omega_{\rm visible}$  could easily 
exist at vastly different scales.

 Another fundamental ratio 
is the baryon to entropy ratio at present time
\be
\label{eta}
\eta\equiv\frac{n_B-n_{\bar{B}}}{n_{\gamma}}\simeq \frac{n_B}{n_{\gamma}}\sim 10^{-10}.
\ee
In our proposal (in contrast with conventional baryogenesis frameworks) this ratio 
is determined by the formation temperature $T_{\rm form}\simeq 41 $~MeV  at which the nuggets and 
antinuggets complete their formation.  We note that 
$T_{\rm form}\sim \Lambda_{\rm QCD}$. This temperature    is determined by the observed  ratio (\ref{eta}). The $T_{\rm form}$  assumes a typical QCD value, as it should as there are no any small parameters in QCD, see Fig. \ref{phase_diagram}.

One should add here that the numerical smallness of the factor (\ref{eta}) in the AQN framework is not due to some small parameters 
which are normally introduced  in the WIMP 
(Weakly Interacting Massive Particles)-based  proposals on baryogenesis. Instead, this small factor  is a  result of an exponential sensitivity of (\ref{eta}) to the temperature as $\eta\sim \exp (- {m_p}/{T_{\rm form}})$ with   the proton's mass being  numerically  large factor when $m_p$  is written in terms of the QCD critical temperature $m_p\simeq 5.5 T_c$ with  $T_c\simeq  \Lambda_{\rm QCD}$. 

To reiterate the same claim: all factors  entering the expression for $\eta$  within AQN framework are the QCD originated parameters. Exponential sensitivity to these parameters generates numerically  small ratio (\ref{eta}) we observe today. 
 
Unlike conventional dark matter candidates, such as WIMPs 
   the dark-matter/antimatter
nuggets are strongly interacting and macroscopically large
 nuclear density objects with a typical size $(10^{-5}-10^{-4})$ cm, and  the baryon charge which 
ranges from $B\sim 10^{23}$ to $B\sim 10^{28}$. However, they do not contradict to any of the many known observational
constraints on dark matter or
antimatter  for three main reasons~\cite{Zhitnitsky:2006vt}:\\
  	{\bf 1.} They carry very large baryon charge $|B|>10^{23}$ which is determined by the size of the nugget $\sim m_a^{-1}$. As a result, 
	  the number density of the nuggets is very small $\sim B^{-1}$. Therefore, their 
non-gravitational  interaction with visible matter is  highly suppressed
and they  do not destroy conventional picture for the structure formation  and cosmic microwave background (CMB)  fluctuations;\\
	{\bf 2.} The nuggets has a huge mass $M_{\rm nugget}\sim m_pB$, therefore the effective interaction is very small $\sigma/M_{\rm nugget}\sim10^{-10}{\rm cm}^2/{\rm g}$, which is evidently well below the upper limit of the conventional DM constraint $\sigma/M_{\rm DM}<1{\rm cm}^2/{\rm g}$. This is the main reason why the AQN behave as a cold DM  from the cosmological point of view; \\
	{\bf 3.} The quark nuggets have very large binding energy due to the large gap $\Delta\sim100$ MeV in the CS phase. Therefore, 
in normal circumstances 
the strongly bound baryon charge is unavailable to participate in the big bang nucleosynthesis (BBN) at $T\approx1{\rm MeV}$, long after the nuggets had been formed.
 
We emphasize that the weakness of the visible-dark matter interaction 
in this model is due to a  small geometrical parameter $\sigma/M \sim B^{-1/3}$ 
  which replaces 
the conventional requirement of sufficiently weak interactions for WIMPs. 
While all interaction  effects  are expected to be, in general, strongly  suppressed due to these features, still  a number of  interesting observable 
phenomena due to the AQN interaction with visible matter may occur as a result of some specific enhancement mechanisms. 

  In particular,  it is known  that the  spectrum from galactic center (where the dark and visible matter densities assume the high   values)
contains several excesses of diffuse emission the origin of which 
is unknown, the best 
known example being the strong galactic 511~KeV line. If the nuggets have the  average  baryon 
number in the $\langle B\rangle \sim (10^{25}-10^{26})$ range they could offer a 
potential explanation for several of 
these diffuse components.  
\exclude{(including 511 KeV line and accompanied   continuum of $\gamma$ rays in 100 KeV and few  MeV ranges, 
as well as x-rays,  and radio frequency bands). }
It is important to emphasize that a comparison between   emissions with drastically different frequencies in such  computations 
 is possible because the rate of annihilation events (between visible matter and antimatter DM nuggets) is proportional to 
one and the same product    of the local visible and DM distributions at the annihilation site. 
The observed fluxes for different emissions thus depend through one and the same line-of-sight integral 
\be
\label{flux1}
\Phi \sim R^2\int d\Omega dl [n_{\rm visible}(l)\cdot n_{DM}(l)],
\ee
where $R\sim B^{1/3}$ is a typical size of the nugget which determines the effective cross section of interaction between DM and visible matter. As $n_{DM}\sim B^{-1}$ the effective interaction is strongly suppressed $\sim B^{-1/3}$. The parameter $\la B\ra\sim (10^{25}-10^{26})$  was fixed in this  proposal by assuming that this mechanism  saturates the observed  511 KeV line   \cite{Oaknin:2004mn, Zhitnitsky:2006tu}, which resulted from annihilation of the electrons from visible matter and positrons from antinuggets.   Other emissions from different frequency bands  are expressed in terms of the same integral (\ref{flux1}), and therefore, the  relative  intensities  are unambiguously and completely determined by internal structure of the nuggets which is described by conventional nuclear physics and basic QED. In particular, this model    offers a 
potential explanation for several of 
these diffuse components (including 511 KeV line and accompanied   continuum of $\gamma$ rays in 100 KeV and few  MeV ranges, 
as well as x-rays,  and radio frequency bands). For further details see the original works 
\cite{Oaknin:2004mn, Zhitnitsky:2006tu,Forbes:2006ba, Lawson:2007kp,
Forbes:2008uf,Forbes:2009wg,Lawson:2012zu}   with specific computations in different frequency bands in galactic radiation, and a short overview
\cite{Lawson:2013bya}.  

Another domain where the coupling between AQN and visible matter could produce some observable effects is related to  the recent EDGES (Experiment to Detect the Global Epoch of reionization Signatures) observation of a stronger than 
anticipated 21 cm absorption  \cite{Bowman:2018yin}. It has been argued in  \cite{Lawson:2018qkc} that this stronger than anticipated  21 cm absorption  can find its natural  explanation within the AQN framework. 
  The basic idea is that the extra   thermal 
emission from AQN  dark matter at early times  produces the required intensity (without  adjusting  any parameters) to explain the recent EDGES observation.

   Yet another the AQN-related effect might be intimately linked to the so-called ``solar corona heating mystery".
  The  renowned  (since 1939)  puzzle  is  that the corona has a temperature  
$T\simeq 10^6$K which is 100 times hotter than the surface temperature of the Sun, and 
conventional astrophysical sources fail to explain the extreme UV (EUV) and soft x ray radiation 
from the corona 2000 km  above the photosphere. Our comment here is that this puzzle  might  find its  
natural resolution within the AQN framework as recently argued in 
 \cite{Zhitnitsky:2017rop,Raza:2018gpb}. 
 
To be more specific,   if one estimates   
the extra energy being injected   when the 
anti-nuggets annihilate with the solar material  one obtains
a total extra energy $\sim 10^{27}{\rm erg}/{\rm  s}$    which 
automatically  reproduces   the   observed EUV and soft x-ray energetics  \cite{Zhitnitsky:2017rop}. 
This estimate is derived  exclusively in terms of known  dark matter density 
$\rho_{\rm DM} \sim 0.3~ {\rm GeV cm^{-3}}$ and dark matter  velocity 
$v_{\rm DM}\sim 10^{-3}c $ surrounding the Sun  without adjusting any  parameters 
of the model. This estimate is strongly supported by Monte Carlo numerical computations  
\cite{Raza:2018gpb} which suggest that most annihilation events occur precisely at the 
so-called transition region at an altitude of $2000$ km, where is known that drastic changes 
in temperature and density occur.  

In the AQN framework the  baryon number distribution 
must be in the range
\be
\label{B-range-flares}
 10^{23}\leq |B|\leq 10^{28}
 \ee
 to be consistent with modelling on solar corona heating related to the energy injection events (the so-called ``nanoflares")
 with typical energies $E_{\rm nanoflares}\sim (m_pc^2) B$. 
It is a highly nontrivial consistency check 
for the proposal \cite{Zhitnitsky:2017rop, Raza:2018gpb} that the required window   
(\ref{B-range-flares}) for nanoflares  is consistent with the range of mean baryon number allowed 
by the  axion and dark matter search constraints as these 
come from a number of different and independent  constraints extracted from astrophysical, 
cosmological, satellite and ground based observations.

A ``smoking gun" supporting the proposal  \cite{Zhitnitsky:2017rop,Raza:2018gpb} on the nature of the EUV  from corona 
would be the observation of the  axions which will be radiated from the corona when  the nuggets get disintegrated in the Sun.
The corresponding computations have been carried out recently in  \cite{Fischer:2018niu,Liang:2018ecs}.  Presently the CAST   (CERN Axion Search Telescope) Collaboration  has taken a significant step to upgrade the instrument to make it    sensitive to the spectral features of the axions produced due to  the  AQN annihilation events  on   the Sun.

Another inspiring observation indirectly supporting the AQN scenario  can be explained as follows.
 It was recently claimed in ref. \cite{Zioutas}   that a number of highly unusual phenomena   observed in the solar atmosphere can be explained by  the gravitational lensing of ``invisible" streaming matter towards the Sun. The phenomena include,  but not limited to  such irradiation as the EUV emission, frequency of X and M flare occurrences, etc. 
Naively, one should not  expect    any correlations between the   flare occurrences, the intensity of the EUV radiation,    and the position of the planets. 
Nevertheless,     the analysis  \cite{Zioutas} obviously demonstrates that this naive expectation is not quite correct. At the same time,
the emergence of  such correlations  within AQN framework is a quite  natural effect. This is because the dark matter AQNs can play the role of the ``invisible" matter in ref. \cite{Zioutas}, which triggers otherwise unexpected solar activity sparking also the large flares \cite{Zhitnitsky:2018mav}. Therefore, the observation of the correlation between the EUV intensity and frequency of the flares  can be considered as an additional supporting argument  of  the AQN related dark matter explanation of the observed EUV irradiation  
because both effects are originated from the same dark matter AQNs.  

We conclude this overview section on the AQN model with the following comment. 
The AQN framework   is consistent with all known astrophysical, cosmological, satellite and ground based constraints. Furthermore, in a number of cases (when   an enhancement factors have emerged) some   observables become very close to present day constraints. 
In fact, in some cases the predictions of the model  may explain a number of the  long standing mysteries as highlighted above. 

The goal of the present work  is to argue   that there is one more such case
when some   enhancements (during the BBN   times) may  lead to   the   consequences which are observed today.  
 To be specific, the claim of the present work is that  the the  nuclei with $Z\geq 3$ during  (or soon after BBN times) 
 are prone to be trapped by negatively charged  antinuggets   at $T\simeq 20$ KeV. Some portion of these  $Z\geq 3$ nuclei will be eventually  annihilated inside the cores of the antinuggets. 
The probability for this  process to occur will be proportional to the  enhancement factor $\sim \exp(Z)$ which, as we argue below, will overcome  a generic feature that    the nuggets play no role in  BBN physics   as stated in  item {\bf 3} above.   
 If further analysis and studies  confirm  this claim it would  represent a long-awaited   resolution of the Primordial Lithium Puzzle within AQN framework.

\section{The internal structure of the nuggets}\label{sec:electrosphere}
In this section we overview some essential technical details related to the internal structure of the AQN.
We start in subsection \ref{pre-BBN} with overview of key points from \cite{Lawson:2018qkc} with analysis of the AQNs at high temperature $T> 2m_e$ when large number of positrons are   present in the system.   In Subsection \ref{sec:electr-struct}
we overview few important results from \cite{Forbes:2009wg} on structure of electro-sphere at low temperature $T\sim $eV.
These results will play an important  role in our main section \ref{sec:suppression} where we present  a precise mechanism which is capable to   suppress the production of nuclei with $Z\geq 3$, which represents the main goal of our studies.

\subsection{Pre-BBN cosmology: AQN annihilation events  and energy injection}\label{pre-BBN} 
We follow  \cite{Lawson:2018qkc} to highlight  few estimates related to the AQN evolution    during the pre-BBN cosmology
with $T> 2 m_e\sim 1$ MeV when the   number  densities of electrons, positrons, baryons  and photons can be  estimated as follows
\be
\label{n_B}
n_B&\sim& n_{\gamma}\eta, ~~ \eta\sim 5\cdot 10^{-10}, ~~ n_{\gamma}\sim \frac{2}{\pi^2} T^3,  
\\
  n_e&\sim& n_{e^+}\sim n_{\gamma}, ~~~ n_B\sim  10^{22} {\rm cm^{-3}} \left(\frac{T}{1~ MeV}\right)^3.  \nonumber 
\ee
 as the thermodynamical equilibrium is maintained in surrounding plasma. 
As we highlight  below,  the presence of the AQNs  does not modify the thermodynamics of the plasma at high temperature
and relations (\ref{n_B}) basically stay the same. In particular, the  rate of energy injection is negligible in comparison with average plasma energy density. Therefore, 
the presence of the nuggets    does not modify the standard  pre-BBN cosmology.

The basic reason for this  conclusion as mentioned in item ${\bf 1}$  from  previous section is that the number density of the nuggets $n_{\rm AQN} \sim B^{-1}$ is 
very tiny due to the very  large baryon charge of a nugget, $B> 10^{23}$. A small energy injection rate to be estimated below is the direct manifestation of this suppression  factor. 

The number of $e^+ e^-$ annihilation events per unit time between electrons from plasma and positrons from nugget's electrosphere for a single nugget can be estimated as follows 
\be
\label{1}
\frac{dN}{dt}\sim 4\pi R^2 n_{\rm e} c
\ee
where  $n_e$ is number density of electrons from plasma, and we assume that electrons hitting the nugget of size $R$ will get annihilated as the density of the positron in electrosphere is large as we discuss in next subsection \ref{sec:electr-struct}.
The energy injection per unit time for a single nugget can be estimated from (\ref{1}) by multiplying a typical  energy $\sim \mu_{e^+}$ being produced as a result of annihilation:
\be
\label{2}
\mu_{e^+}\frac{dN}{dt}\sim 4\pi R^2  n_{\rm e} c\mu_{e^+},.
\ee
where $\mu_{e^+}$ is the chemical potential of the positrons from nugget's electrosphere, see next subsection. 
The energy injection per unit volume per unit time can be estimated as
\be
\label{3}
\frac{dE}{dVdt}\sim 4\pi R^2 n_{\rm AQN} n_{\rm e}c \mu_{e^+}, 
\ee
where the AQN number density $n_{\rm AQN} $ from (\ref{3}) 
 is estimated as follows 
\be
\label{4} 
n_{\rm AQN}\sim \frac{n_B}{\la B\ra} \sim \eta \frac{n_{\gamma}}{\la B\ra} \sim \frac{T^3 \eta}{\la B\ra}\sim 5\cdot 10^{-36}T^3 ,
\ee
where $ \eta$ is conventional the baryon to photon ratio factor (\ref{n_B}). For numerical estimates  we used $\la B\ra \simeq 10^{26}$. We want to estimate the total amount of energy being injected into the system per unit volume  during  mean free time $\tau\sim (\alpha^2T)^{-1}$ which is defined as a typical time between collisions. During time $\tau$ the system adjusts any injection of the external energy into the system due to the  fast equilibration. 
We want to compare this extra injected energy (as a result of annihilation events of electrons with positrons from electrosphere) with typical energy density in the system $\sim (T n_{\rm e})$, i.e.
we consider the dimensionless ratio
\be
\label{5}
\frac{\tau}{T n_{\rm e}}\cdot\left(\frac{dE}{dVdt}\right)\sim    \frac{\left(\eta R^2 T \mu_{e^+}\right)}{\alpha^2\la B\ra}\sim 10^{-18}\left(\frac{T}{1~ \rm MeV}\right), ~~~~
\ee
where we use $\mu_{e^+}\sim 10$ MeV for numerical estimates, see next subsection.

It is clear that such small amount of energy injected into the system will be quickly equilibrated within the system such the standard pre-BBN cosmology remains intact.  In other words, the conventional equation of state, and conventional evolution of the system is unaffected  by presence of AQNs at $T> 2 m_e$ when the  positrons in plasma are in thermal  equilibrium and easily available to 
replace the positrons from antinugget's electrosphere.  The basic reason for this conclusion is   due to the fact that the number density
of the AQNs is very tiny according to (\ref{4}) such that the conventional  interaction of the AQNs with surrounding material in normal circumstances is strongly suppressed due to $\la B\ra^{-1}$ factor.

\subsection{Electrosphere Structure}\label{sec:electr-struct}
We need one more ingredient for our future analysis in Sect.  \ref{sec:suppression}
suggesting that in some circumstances the AQNs can drastically modify 
the conventional BBN outcome. This additional  ingredient is related to analysis of electrosphere
of the AQN at $T \ll m_e$ when it is placed into a dilute system (such as Inter-Stellar Medium (\ISM)) where very few  particles are present in the system.   The corresponding studies have been carried out in  \cite{Forbes:2009wg}
and played an important role in analysis of   access of radiation from the galactic center
as reviewed in previous section, see few paragraphs around eq.(\ref{flux1}). 
We overview the basic ideas of computations \cite{Forbes:2009wg} in this subsection  to make our presentation self-contained.

The basic idea of  \cite{Forbes:2009wg} is to use 
Thomas-Fermi analysis including the full relativistic electron
equation of state required to model the relativistic regime close to
the nuclear core.  One should  emphasize  that
the presence of electrosphere itself is a very generic phenomenon, and  its main features  
are determined by    the boundary conditions deep inside the nugget (being in CS phase) where
the lepton's chemical potential is fixed as a result of the beta  equilibrium,  similar to 
 analysis of refs~\cite{Alcock:1986hz, Kettner:1994zs} in 
  the context of strange
stars.

The density profile of the electrosphere 
has been derived in  \cite{Forbes:2009wg} 
from a density functional theory   after neglecting the exchange
contribution, which is suppressed by the weak coupling
$\alpha$.  The electrostatic potential $\phi(r)$ must satisfy the
Poisson equation 
\begin{equation}
  \label{eq:Poisson}
  \nabla^2 \phi(r) = -4\pi e n(r).
\end{equation}
where $e n(r)$ is the charge density which can be expressed in terms of  the chemical potential
\begin{equation}
  \label{eq:mu}
  \mu(r) =  \mu - e\phi(r).
\end{equation}
   The
resulting equation  assumes the form
\be
\label{eq:Poisson1}
      &&\nabla^{2}\mu(r) = 4\pi \alpha n[\mu(r)], ~~~
      \epsilon_{p} = \sqrt{p^2+m^2} \\
    &&n[\mu] = 
    2\!\int\!\!\frac{\d^{3}{p}}{(2\pi)^3}\!\left[
      \frac{1}{1+e^{(\epsilon_{p}-\mu)/T}} -
      \frac{1}{1+e^{(\epsilon_{p}+\mu)/T}}\right],\nonumber
 \ee
In~(\ref{eq:Poisson1})  both particle and
antiparticle contributions have been explicitly  included, and  $e^2 = \alpha$  in units where $\hbar = c = 4\pi\epsilon_{0} = 1$.

Few comments on the boundary conditions which have been imposed in analysis \cite{Forbes:2009wg}.
The boundary conditions at $r=R$ (at the nugget' surface)  are determined by the beta  equilibrium, similar to 
 analysis in 
  the context of strange
stars~\cite{Alcock:1986hz, Kettner:1994zs}.  In context of CS dense matter a similar condition applies. 
However, a precise structure in CS phase is not known, and therefore, $\mu (R)$ at the boundary $R$  which depends   on  
equation of state for the quark-matter phase is also   not
known.  Typical QCD based estimates  suggest that lepton chemical potential  $\mu$ is of order $ \approx 25$~MeV, see e.g. review \cite{Alford:2007xm}. This is  the value which  was adopted in \cite{Forbes:2009wg} for numerical estimates.

 Another  parameter from studies  \cite{Forbes:2009wg} is the outer radius $r_{*}$  of electrosphere 
 where nuggets 
will ``radiate'' the
loosely bound positrons  until the electrostatic potential is
comparable to the temperature $\alpha Q/r_{*} \sim T$, where $eQ$ is the charge 
of the AQNs due to the ionization at temperature $T$.

The equation (\ref{eq:Poisson1}) with the corresponding boundary conditions
as explained above has  been solved numerically ~ \cite{Forbes:2009wg}
with the profile function  which smoothly interpolates from ultra-relativistic regime 
deep inside the nugget and one for the non-relativistic Boltzmann regime~ \cite{Forbes:2009wg}
where $n[\mu] \propto e^{\mu/T}$ at $z\equiv (r-R)$ far away from the nugget's  surface which is defined as $z=0$. 

For non-relativistic Boltzmann regime one can approximately describe the density of positrons in electrosphere as 
follows  \cite{Forbes:2008uf}:
 \begin{equation}
 \label{eq:Boltzmann}
  n_{e^+}(z) = \frac{T}{2\pi\alpha}\frac{1}{(z + \bar{z})^2}, ~~~~ \frac{1}{\bar{z}}=m_e \sqrt{2\pi\alpha} \left(\frac{T}{m_e}\right)^{1/4}
\end{equation}
where $\bar{z}$ is the integration constant is chosen to match the Boltzmann regime at sufficiently  large $z\gg \bar{z}$. Numerical 
studies ~ \cite{Forbes:2009wg}  support the approximate analytical  expression (\ref{eq:Boltzmann}).

The same expression   (\ref{eq:Boltzmann}) for the positron density in electrosphere  as the solution of the Thomas-Fermi  equations can be  also represented in terms of the  chemical potential   as these two parameters are related according to eqs.(\ref{eq:mu}) and  (\ref{eq:Poisson1}), see  \cite{Forbes:2008uf} with details, 
\be
\label{potential}
n_{e^+}[\tilde{\mu}]=\sqrt{2}\left(\frac{mT}{\pi}\right)^{3/2} e^{\tilde{\mu}/T}, ~~~~ \tilde{\mu}\equiv \mu(r)-m_e, 
\ee
where we redefined the chemical potential by removing a large constant term $m_e$ to make it appropriate  for non-relativistic regime.
 
We conclude this overview section on electrosphere's structure with the following comments. The results presented above are 
well suited for studies of the AQNs in  low temperature  and low density environment such as
Inter-Stellar Medium 
 in our galaxy. The goal of the present work is drastically different as it deals with relatively high temperature plasma with $T\sim 20$ KeV soon after  the BBN
epoch. Why the temperature $T\sim 20$ KeV  is so special for our analysis? This is because at   $T\approx  20$ KeV the positron number density in plasma assumes  the same order of magnitude as the baryon number density $n_B\simeq \eta n_{\gamma}$.  When the temperature becomes slightly  lower, i.e. $T< 20 $~KeV, the positrons will be soon completely annihilated while the proton number density essentially remains the same  as $n_B= \eta n_{\gamma}$.  In context of our work it implies that 
the screening of the  negative charge of the antinugget will be provided by the protons at  $T< 20 $~KeV as explained in next section.

\section{The AQN-induced suppression mechanism for large $Z\geq 3$ }\label{sec:suppression}
\subsection{Preliminaries} 
The starting point for our analysis is the observation that the high temperature environment leads to ionization of the 
loosely bound positrons such that the antinuggets will become   negative charged ions with  charge  $-Q$  estimated as follows
  \be
  \label{Q}
Q\simeq 4\pi R^2 \int^{\infty}_{z_0}  n(z)dz\sim \frac{4\pi R^2}{2\pi\alpha}\cdot \left(T\sqrt{2 m_e T}\right)~~~~~
  \ee
 where we assume that the positrons with $p^2/(2m_e)< T$ will be stripped off the electrosphere as a result of high temperature $T$. These  loosely bound positrons 
 are localized mostly at outer regions of electrosphere at distances $z> z_0= (2 m_e T)^{-1/2}$ which motivates our cutoff in estimate (\ref{Q}). For   these estimates   we also used an approximation (\ref{eq:Boltzmann}) from Section \ref{sec:electr-struct} to express  $Q$ in a simple analytical form (\ref{Q}).
 
 One should emphasize that the mean-field approximation is not justified at very large distances. Furthermore, an approximate 
 analytical expression (\ref{eq:Boltzmann}) used in estimate (\ref{Q}) is sufficiently good approximation for distances close to the surface  when one-dimensional treatment in terms of $z\ll R$ is appropriate,  but breaks down for distances $z\gg R$ when effectively 3D treatment is required. However, these deficiencies do not drastically modify our estimate (\ref{Q}) for the total charge $Q$ 
 as the dominant contribution to the integral (\ref{Q}) comes from small $z\ll R$ where the approximate solution (\ref{eq:Boltzmann}) is valid. 
 
 The key observation of the present work is that the positrons which are stripped off due to the high temperature will be replaced  by the positively charged protons\footnote{In context of the present work the temperature $T\sim 20$ KeV is very important parameter  just because it  corresponds to the  epoch when the positron plasma density  becomes the same order of magnitude as the baryon density, i.e. $n_B \sim n_{e^+}$.  As a result, the proton's from plasma is capable to   screen the negative electric charge of the antinuggets; for higher temperature this role  is  played by   the positrons which were much more abundant in plasma at $T>20 $ KeV.}. This process inevitably should take place to neutralize  the large negative charge $Q$ estimated by eq (\ref{Q}). Before we proceed with corresponding estimates it is very instructive   to compare the charge density of the protons which will be accumulated  in vicinity of the nugget's surface 
 with average proton's density (\ref{n_B}) far away from the nuggets in plasma.
 
 The proton's charge density will have  essentially the same qualitative behaviour  as   (\ref{eq:Boltzmann})
 similar to  the positron's charge density (if the positrons  were not stripped off) representing  the  solution of the Thomas -Fermi  equation. The only difference is that for proton's density profile $n_{p}(z)$ we fix the integration constant $\bar{z}_p$ assuming that the protons  neutralize the negative charge $Q$ given by (\ref{Q}). This boundary conditions should be contrasted with  our studies in   Section \ref{sec:electr-struct} for positrons 
  where boundary conditions were imposed by matching  the chemical potential generated due to the $\beta$ equilibrium deep inside the nuggets.  
  
  Therefore, 
the estimation for the proton density in close vicinity of the surface   $n_{p}(z)$ is very  similar to estimates  (\ref{eq:Boltzmann}) when the Thomas -Fermi approximation can be trusted, i.e.
\begin{equation}
 \label{eq:proton}
  n_{p}(z) = \frac{T}{2\pi\alpha}\frac{1}{(z + \bar{z}_p)^2}, ~~~ z\ll R
  \end{equation} 
 where the integration constant $\bar{z}_p$ is fixed by the condition
 \be
  \label{z_p}
  \int_0^{\infty}n_p(z)dz=Q=\frac{4\pi R^2T}{2\pi\alpha} \frac{1}{\bar{z_p}}, ~~~  \frac{1}{\bar{z_p}}=\sqrt{2m_e T}.~~~
  \ee
  The density of protons at $z=0$   is huge
\be
 \label{n_0}
 n_p (z=0)=\frac{T}{2\pi\alpha} \frac{1}{\bar{z_p}^2}\simeq  1.3\cdot 10^{30} {\rm cm^{-3}}
 \left(\frac{T}{20~ {\rm KeV}}\right)^2.~~~
 \ee
 One should emphasize that the total accumulated charge $Q$ due to the screening by protons (\ref{z_p}) is the surface effect. Therefore,  baryon charge in form of the protons (\ref{z_p}) represents a very small fraction of the total baryon charge hidden in the nuggets, i.e. $Q/B\ll 1$. The direct annihilation of the protons surrounding the anti-nugget is strongly suppressed due to the large gap in CS phase as mentioned  in item {\bf 3} in Section \ref{sec:AQN}.
 
 It is instructive to   compare (\ref{n_0}) with the protons' density $n_B$ outside the nugget in plasma
 given  in  (\ref{n_B}).
The ratio of the densities $n_B$ and $n_p(z=0) $ is estimated as   
\be
\label{n_B_ratio}
\left[\frac{n_B(T)}{n_p(z=0, T)}\right]\sim 6\cdot 10^{-14} \left(\frac{ T}{20 ~{\rm KeV}}\right).  
\ee
It depends on $T$ because $n_B\sim T^3$ has conventional scaling while $n_p(z=0)\sim T^2$ is the surface effect, and  does not follow the free particle distribution. 

We need yet another ingredient for   presenting the suppression  mechanism for heavy nuclei with $Z\geq 3$.
We define the capture radius $R_{\rm cap}(T)$ as the distance when the external protons from plasma have sufficiently low velocities
such that they can be captured (trapped) by long ranged Coulomb forces and become bounded to the antinugget, i.e.
\be
\label{capture}
\frac{\alpha Q(r)}{r} > \frac{m_p v^2}{2}\approx T ~~~{\rm for}~~~ r\leq R_{\rm cap}(T).
\ee
We estimate $R_{\rm cap}(T)$ (which is obviously much larger than the size of the nugget $R_{\rm cap}(T)\gg R$)  and related parameters in next section \ref{mechanism}. It is expected that at $r\geq R_{\rm cap}$ the density of the protons in electrosphere becomes the same order of magnitude as  the average density of the plasma $n_B$ which itself is determined exclusively by the temperature according to (\ref{n_B}). 

Final  comment we would like to make in this subsection  is that the ratio $\xi$ defined as 
\be
\label{xi}
\xi (r, T)\equiv\frac{\alpha Q(r)}{r T}
\ee
will play very important role in our arguments which follow. The parameter $\xi (r, T)$ obviously describes the ratio of the potential 
 binding energy in comparison with kinetic energy $\sim T$.  Important parameter to be discussed below 
 is the average characteristic $\la \xi (T)\ra$ which represents the mean value  of this ratio over entire ensemble of the particles surrounding the antinugget. 
 \subsection{Few relevant estimates}\label{estimates}
 
   We start our analysis with numerical  estimate  the capture radius $R_{\rm cap}(T)$  as defined by (\ref{capture}).
  We assume   that the density $ n_p(r=R, T)$ has a  power like behaviour at $r\gtrsim  R$ with exponent $p$.
  This assumption is consistent with our numerical studies \cite{Forbes:2009wg} of the electrosphere with $p\simeq 6$. It is also consistent with conventional Thomas-Fermi model at $T=0$, see e.g. Landau textbook  
  \cite{Landau}\footnote{In notations of ref.
   \cite{Landau} the dimensionless function $\chi (x) $ behaves as $\chi\sim x^{-3}$ at large $x$. The  potential $\phi=\chi (x)/x$ behaves as 
  $\phi\sim x^{-4}$. The density of electrons in Thomas-Fermi model scales as $n\sim \phi^{3/2}\sim x^{-6}$ at large $x$.}. 
  We keep parameter $p$ to be arbitrary to demonstrate  that our main claim is not very sensitive to our assumption on numerical value of $p$.

 Therefore,   we parameterize the density as follows
\be
\label{density}
n_p(r, T)\simeq n_p(r=R, T) \left(\frac{R}{r}\right)^p, ~~~
\ee
where $n_p(r=R)\equiv n_p(z=0)$ is the surface density determined by the eq. (\ref{n_0}). We can now  estimate the effective capture distance 
$R_{\rm cap}$. It can be approximately computed from the following condition
\be
\label{R_cap}
n_p(R_{\rm cap}, T)\simeq n_p(z=0, T) \left(\frac{R}{R_{\rm cap}}\right)^p\simeq n_B(T)
\ee
where $n_B (T)$ defined by (\ref{n_B})  is the average  proton number density far away from the nuggets. 
From (\ref{R_cap}) and (\ref{n_B_ratio}) we estimate effective capture distance 
$R_{\rm cap}$ as follows
\be
\label{R_cap1}
  \left(\frac{R}{R_{\rm cap}}\right)^p&\simeq& 6\cdot 10^{-14}\left(\frac{T}{20~ \rm KeV}\right) 
  \ee
In particular for $p\simeq 6$ the   effective capture distance 
$R_{\rm cap}$ is of order
\be
\label{R_cap2}
   R_{\rm cap} \simeq  1.6\cdot 10^2\cdot \left(\frac{20~ \rm  KeV }{T}\right)^{\frac{1}{p}} R\simeq 3 \cdot 10^{-3} {\rm cm}  
\ee
for typical size of the  nugget $R\simeq 2\cdot 10^{-5}{\rm cm}$ and  $p=6$.

Our next task is to estimate the screened charge $Q(r\sim R_{\rm cap})$ at large distances $r\sim R_{\rm cap}$  far away from the nugget's core.  We assume that this is the region where the power like behaviour (\ref{density}) for the density $n_p (r)$
still holds, and the expected exponential tail (which    cannot be accommodated within a simple mean-field approximation 
  adopted in this work)  is  not  yet operational.  This screened charge is obviously must be much smaller that the original charge (\ref{z_p}). Indeed, within our framework one can compute the screened charge by integrating from $R_{\rm cap}$ to infinity 
  instead of accounting for the cancellations between the original negative charge of the antinugget and positive   charge of the surrounding protons, i.e.
   \be
 \label{Q_screened} 
   Q(R_{\rm cap})   \simeq \int_{R_{\rm cap}}^{\infty} 4\pi r^2dr n_p(r) \sim \frac{4\pi n_B(T)R^3_{\rm cap}(T)}{(p-3)},~~~~~
 \ee
 where $n_p(r)$ is the charge density determined by   (\ref{density}), and we expressed the final formula in terms of the background 
baryon density $n_B(T)$ at temperature $T$.  
  It is known that at much larger distances the behaviour must be changed to $\exp(-r)$ due the screening at very large distance  $r$, but the integral  (\ref{Q_screened}) is saturated by much smaller $r\sim R_{\rm cap}$; therefore we ignore the small corrections due to the exponential tail. 
\exclude{
Another related comment is as follows. The charge $Q$ entering (\ref{enhancement})  is to be computed at point $r$ is a result of very strong cancellation between the bare negative charge $Q_0$ rom the nuggets and high density positrons and protons close to the nugget's surface (\ref{n_p}). Numerically, the result of this strong cancellation is easy to estimate by computing the rest of the charge by taking the integral from $R_{\rm cap}$ to $\infty$. This procedure assumes that the total charge of the system will be eventually screened at very large distances.
}
It is useful for what follows  to represent formula (\ref{Q_screened}) 
in the form 
which explicitly shows the $T$ dependence and the algebraic exponent  $p$:
  \be
 \label{Q_screened1} 
   Q(R_{\rm cap})   \sim   10^{10}\cdot  \left(\frac{T}{20~ \rm KeV }\right)^{3(1-\frac{1}{p})},
 \ee
where for the numerical estimates we use $R\simeq 2\cdot 10^{-5}{\rm cm}$.  

It is also instructive to estimate the number of  particles being  affected by the presence of the AQNs in the system. To estimate this ratio of the ``affected particles" one should multiply (\ref{Q_screened1}) 
by  the density of the antinuggets (\ref{4}) and compare the obtained result  with the average baryon density $n_B(T)$ in plasma, i.e.
\be
\label{delta}
\frac{\delta n_p}{n_p} \sim \frac{\left[n_{\rm AQN}\cdot  Q(R_{\rm cap})\right] }{n_B}\sim   3\cdot  10^{-16}
\left(\frac{T}{20~ \rm KeV }\right)^{3(1-\frac{1}{p})}.~~~~
\ee
The  density of the antinuggets $n_{\rm AQN}$ in this formula  is estimated as 
\be
\label{n_AQN}
n_{\rm AQN}\simeq \frac{n_B}{\langle B\rangle}\cdot 5\cdot \frac{3}{5},
\ee
where factor $3/5$ accounts for the portion  of the antinuggets, while factor $5$ accounts for approximate ratio $\rho_{\rm DM}\simeq 5 \rho_{\rm B}$ when  it is  assumed  that the DM is saturated by nuggets and antinuggets. 

The number of affected particles ${\delta n_p}/{n_p}$  as one can see from estimate (\ref{delta})  is absolutely negligible, as expected. This claim is similar to analogous estimates in pre-BBN cosmology expressed by formula (\ref{5}). In both cases the strong  suppression  is  a result of very  tiny  number density
of the AQNs   such that the conventional  interaction of the AQNs with surrounding material in normal circumstances is strongly suppressed by factor  $\la B\ra^{-1}$   in comparison with visible   baryon interactions.

 \subsection{Suppression Mechanism for heavy nuclei}\label{mechanism}
We are in position now to formulate the basic idea for the suppression mechanism for heavy nuclei which goes as follows. 
In previous sections in our analysis related to  the screening of the original antinugget's charge   we had assumed that all the particles which screen the  negative electric charge $-e Q$  are the protons which have positive unit electric charge $+e$.
The corresponding density $n_p$ in the vicinity of the nugget's core is determined by eq. (\ref{eq:proton}), and represents the self-consistent solution in mean-field approximation. The presence of 
heavier nuclei with $Z> 1$ do not qualitatively change  the structure of the electrosphere as long as densities of these nuclei 
$n_Z$ are sufficiently small in comparison with the background proton density $n_B$, i.e. $n_Z\ll n_B$. 

However, the interaction of these heavy nuclei  with the nugget's charge $Q$ is exponentially stronger 
due to the Boltzmann enhancement. It can be seen explicitly from formula (\ref{potential}) represented in terms of the chemical potential which itself is expressed in terms of  electrostatic potential (\ref{eq:mu}). 
We would like to represent the corresponding enhancement factor  in the following way
\be
\label{Z}
 \sim \exp\left[  {\frac{(Z-1)\alpha  Q(r)}{rT}} \right]
\ee 
where we inserted an additional factor $(Z-1)$ into the expression to avoid the double counting.  
Indeed, the protons with  $Z=1$   have been accounted for in the Thomas-Fermi computations leading to (\ref{potential}). Precisely this Boltzmann (Fermi for degenerate case) distribution leads to high density of protons close to the nugget's surface (\ref{n_0}). 

Now we are in position   to 
  compute the relative number of the trapped and captured ions with $Z>1$.
  The corresponding estimate goes exactly in the same way as our estimates with protons (\ref{delta}) with $Q(R_{\rm cap})$ given by (\ref{Q_screened})
\be
\label{captured}
\frac{\delta n_Z}{n_Z} \simeq\frac{4\pi R^3_{\rm cap}}{3}\cdot n_{AQN}\cdot \exp\left[ {\frac{(Z-1)\alpha  Q(r)}{rT}} \right],
\ee
where we inserted  the enhancement   factor (\ref{Z}) as explained above. 
We want to avoid the double counting of the particles with charges $Z=1$. Therefore,   we    introduce   the  factor $(Z-1)$ in (\ref{captured}) and treat it as an enhancement factor for  ions (He, Li and Be) in comparison with protons. 
Note that the density   $n_Z$ is very small in comparison with the density of protons and does not perturb their distribution. In the relative estimate in eq.  (\ref{captured}) it enters the numerator and denominator and cancels out while enhancement factor (\ref{Z}) obviously stays.

  It is convenient to estimate  the dimensionless suppression factor (first two factors from  eq. (\ref{captured}))  as follows
\be
\label{suppression}
  \left[\frac{4\pi R_{\rm cap}^3}{3}\cdot n_{AQN}\right]
\sim  2.7\cdot 10^{-16}
\left(\frac{T}{20~ \rm KeV }\right)^{3(1-\frac{1}{p})}, ~~~~
\ee
  We want to rewrite this dimensionless factor in  the exponential form as all elements are highly (exponentially)  sensitive to many unknown parameters, i.e.
 \be
\label{suppression2}
\left[\frac{4\pi R^3_{\rm cap}}{3}\cdot n_{AQN}\right]\simeq \exp(-X_{\rm Sup}), ~~~ \nonumber \\
 X_{\rm Supp}\simeq 35.8-3\left(1-\frac{1}{p}\right)\ln  \left(\frac{T}{\rm 20~ KeV}\right).
 \ee

 Now we want to argue that the last dimensionless factor $\sim \exp (...)$ in eq. (\ref{captured}) represents a huge enhancement for   ions such as Li with $Z=3$ and Be with $Z=4$, while it  remains to be relatively small for   He with $Z=2$ and vanishes for H with $Z=1$.   
 We proceed with estimates of the enhancement factor entering (\ref{captured}) by assuming  that the screened charge of the antinugget $Q(R_{\rm cap})$ is estimated at $r\simeq R_{\rm cap}$ as given by  in eqs. (\ref{Q_screened}) and  (\ref{Q_screened1}):
  \be
 \label{enhancement2} 
  \left( {\frac{\alpha  Q(R_{\rm cap})}{R_{\rm cap} T}}\right)   \sim 20\cdot \left(\frac{T}{20~ \rm KeV}\right)^{2(1-\frac{1}{p})}.
 \ee
 We want represent the enhancement factor entering (\ref{enhancement2}) in the same exponential way as the suppression factor (\ref{suppression2}), i.e.
 \be
 \label{enhancement3} 
 \exp \left[(Z-1)\cdot  {\frac{\alpha  Q(R_{\rm cap})}{R_{\rm cap} T}}\right]  = \exp(+X_{\rm Enh}), \nonumber \\
 ~~~ X_{\rm Enh}\simeq 20 (Z-1) \cdot \left(\frac{T}{20~ \rm KeV}\right)^{2(1-\frac{1}{p})}.
 \ee
The relative number of the trapped and captured ions defined by   (\ref{captured}) is estimated now as follows 
\be
\label{final}
\frac{\delta n_Z}{n_Z} \simeq e^{(-X_{\rm Supp}+ X_{\rm Enh})}  
\ee
where $X_{\rm Supp}$  and $X_{\rm Enh}$ are estimated by eqs.  (\ref{suppression2}) and (\ref{enhancement3}) correspondingly.  
For our purposes of order of magnitude estimate one can neglect the $\ln$ term in    (\ref{suppression2}) and $1/p$ in (\ref{enhancement3}) to  approximate the final formula for the exponent for $Z=3$ as follows 
\be
\label{final1}
 \left(-X_{\rm Supp}+ X_{\rm Enh}\right)\simeq \left[ -36+40  \cdot \left(\frac{T}{20~ \rm KeV}\right)^{2}\right], 
\ee
  which suggests that relative number of the 
  remaining   Li ions might  be strongly 
 depleted as $ {\delta n({Li}})/{n({Li})}\sim 1$ because  the depletion becomes order of one effect for   $T\approx  20$ KeV.   For Be ions the depletion effect is even stronger. The depletion effect  for He with $Z=2$  can be ignored as the enhancement factor (\ref{enhancement3})  is insufficient 
   to overcome the suppression factor   (\ref{suppression2}) in this case. One should also remark here that we do not discriminate 
   $^6$Li and   $^7$Li in our estimates. In fact in our effective mean field approximation it would be very hard to do so, especially 
   due to the fact that $^6$Li density is strongly suppressed in comparison with $^7$Li. Therefore, the only  claim one can make is that the ions with charges $Z=3$ are strongly depleted as eq. (\ref{final1}) states. 
 
  In our estimates  (\ref{final}), (\ref{final1}) we, of course, assume that the finite portion of the Li ions will be affected by the nuggets during the   cosmic  time $t_0\sim 2\cdot 10^3$s corresponding to the temperature $T\approx 20 $ keV. 
  We refer to Appendix \ref{flux} where we estimate the fluxes of the ions entering the AQNs vicinity of size $R_{\rm cap}$.
  We argue in Appendix \ref{flux} that  a finite portion of all  ions  in entire volume   will be affected by the nuggets.  However, the eventual effect of this impact of the nuggets  is negligible for   light nuclei with $Z\leq 2$ and becomes crucial for heavy   ions with $Z\geq 3$    as our estimates  (\ref{final}), (\ref{final1}) suggest. 
   
  Formula (\ref{final1})  is indeed an amazing result which might be the resolution of the primordial Li problem
 as finite portion of the produced Li gets captured by  the antinuggets at $T\simeq 20~ {\rm KeV}$ soon after the BBN ended. 
 It is important  to emphasize that no any special fitting procedures  have been employed in the estimates presented above.  All parameters which have been used  to arrive to the final expression (\ref{final}), (\ref{final1}) 
assume the same values similar to our  previous studies related to the galactic excess emission, estimates related to EDGES observations, and resolution of the  solar corona heating puzzle within AQN model as reviewed in Sect. \ref{sec:AQN}. 
 
 Few comments are in order. First of all, our assumption on specific algebraic exponent $p\simeq 6$ is not crucial  as the final results are not very sensitive to this assumption. Our estimates are obviously very sensitive to the parameters of the nuggets, such as typical baryon charge $B$, radius of the nugget, $R$, etc. One should emphasize that all these typical parameters are consistent 
 with cosmological, astrophysical and ground based constraints as overviewed in Section  \ref{sec:AQN}. Furthermore, these parameters are consistent with axion search experiments constraints as parameters $R$ and $m_a$ are not independent, but related to each other.

\section{Conclusion and future directions\label{conclusion}}
The main result of the present work is represented by equations (\ref{final}) and (\ref{final1}) which show that the primordial abundance of $^7$Li nuclei could be much smaller than conventional computations  \cite{ParticleData} predict.  The effect is  due to the capture and subsequent  annihilation of Li and Be ions by antinuggets within AQN paradigm.

 A proper 
 procedure would be, of course, integrating over time evolution and averaging over the nugget's size distribution, similar to studies 
 \cite{Raza:2018gpb} on the solar nanoflare distribution as overviewed in 
Section  \ref{sec:AQN}. It was not the goal of the present work to carry out precise computations of the effect. 
Such a study would not be sufficiently  precise procedure anyway   because of a huge (exponential) sensitivity 
to the parameters $R, B$ and their distributions which are 
  not well  known\footnote{A more precise estimate is very hard to carry out as the final result is exponentially sensitive to the detail properties of the nuggets. Indeed, the effect becomes of order   one as a result of cancellation of two very large numbers as one can see from  (\ref{final}), (\ref{final1}).  }.
Rather, our intention  was to demonstrate   that the resolution of the primordial Li puzzle might be a very natural outcome of presence of the AQNs 
in the plasma, and their interactions with the visible matter  soon after BBN formation epoch. 

One should emphasize that the estimate  (\ref{final}), (\ref{final1})  formally makes sense as long as $({\delta n_Z}/{n_Z})\ll 1$.
 However, the point is that  $({\delta n_Z}/{n_Z})$ could  easily assume a value of order  one. It unambiguously implies that a finite portion of Li  and Be nuclei from plasma get trapped by the  antinugget's electrosphere. 

We believe that our order of magnitude estimates (\ref{final}), (\ref{final1})  represent  sufficiently convincing arguments   supporting the claim that  $({\delta n_Z}/{n_Z})\sim 1$ for Li with $Z=3$ and Be with $Z=4$.   Indeed, we demonstrated that the internal properties  of the nuggets are such that the heavy nuclei with $Z\geq 3$ are 
strongly attracted to the antinuggets due to the long range Coulomb forces. These heavy nuclei will be bounded to the antinuggets and will eventually get annihilated in the AQN's cores in subsequent  time evolution 
\footnote{\label{lines}Note that the abundance of Li atoms and ions  is measured using the intensity of their atomic spectral lines. The captured Li nuclei  do not produce atomic spectra so they can not contribute to the measured Li abundance even before their annihilation.}.

One should emphasize that 
the corresponding key parameters of the nuggets which have been used in our estimates have not been specifically chosen 
for the   purposes of the present work devoted  to the resolution of the primordial Li puzzle (as it is normally done in a typical proposal on resolving Li puzzle  within WIMP framework). 
  Instead, all the key parameters have been originally fixed    for very different purposes to satisfy a variety of  constraints from a number of  unrelated  experiments and observations as reviewed in Sect.  \ref{sec:AQN}. 
  
  Therefore, the resolution of the Li puzzle within AQN framework  represents yet another indirect support for  this new paradigm on the nature of DM and baryon charge separation replacing the conventional ``baryogenesis".   The list of these indirect evidences supporting the AQN framework includes (but not limited) such long standing problems as a natural explanation of the observed ratio $\Omega_{\rm dark} \sim   \Omega_{\rm visible}$, renowned     puzzle coined as the ``solar corona heating mystery", recent EDGES observations, to name just a few,  see overview   in Sect.  \ref{sec:AQN}. 
  
 To reiterate:  this AQN model was invented  to explain   the observed ratio (\ref{Omega}) in a natural way as both types of matter (visible and dark) are formed during the same QCD epoch in early Universe, and proportional to one and the same dimensional parameter $\Lambda_{\rm QCD}$.    Precisely the same generic feature plays a key role in the suppression mechanism 
 for  the abundance of  heavy nuclei with $Z>2$ as presented in Sect. \ref{mechanism} because  the visible nuclei and antinuggets made from the {\it same} Standard Model quarks and gluons (but in different phases, the hadronic phase  and CS phase correspondingly).   

Can we study any traces  of  the captured (after BBN epoch) heavy nuclei by antinuggets   today? It is very unlikely as the captured   heavy nuclei will eventually get annihilated in the antinugget's core.  Furthermore, it is hard to expect any specific  electromagnetic signatures as a result of these annihilation processes of the heavy nuclei, see also footnote \ref{lines} with related comments. 

\exclude{
Another possible trace of  the captured  heavy nuclei by antinuggets which may potentially lead to some observable effects is related to some  lose of local "inhomogeneity" of the Universe  on the scales of order the distance between the nuggets.
The corresponding typical distance at present time can be estimated as  $d_{\rm dist}\sim (n_{\rm AQN})^{-1/3}$. The corresponding nugget's density  at present time   can be estimated
as $n_{\rm AQN}\sim \frac{3}{5}\frac{\rho_{DM}}{\la B\ra m_p}$, similar to (\ref{n_AQN}), such that a typical distance between the nuggets is around $d_{\rm dist}\sim 10^9 ~{\rm cm}\sim 10^4 ~{\rm km}$ if one uses the same parameter $\la B\ra\sim 10^{26}$ as discussed in the text. It is very unlikely that such local inhomogeneities of primordial Li ions can be observed on such small scales. 
}

In some circumstances, though,  the antinuggets can be completely disintegrated, for example in the solar corona leading to the extreme UV radiation as reviewed in Sect.  \ref{sec:AQN}. When the AQNs propagate in the earth's atmosphere  they obviously produce some observable effects.  In fact, the propagating of the AQN in 
earth's atmosphere can mimic the  ultra high energy cosmic ray air showers \footnote{It is interesting to note that Antarctic Impulsive Transient Antenna (ANITA)  has  recently observed   upward traveling, radio-detected cosmic-ray-like events with characteristics
closely matching an extensive air showers but with the inverse direction of the shower's cone \cite{ANITA}. 
Furthermore, some strong arguments have been recently presented in  \cite{Fox:2018syq} suggesting that these anomalous events cannot be explained within the SM particle physics, and, therefore,  should be treated as ``dramatic and highly credible evidence of the first new bona fide BSM phenomenon since the discoveries of neutrino oscillations, dark matter, and dark energy".  It is tempting to identify  these anomalous events  with AQNs travelling in deep underground and exiting with angles $27^o$ and $35^o$ as recorded by ANITA.}  as argued in  \cite{Lawson:2010uz,Lawson:2012vk,Lawson:2013bya}.

 When the AQNs travelling in deep earth's  underground it is very unlikely to observe  any specific $E\&M$ signatures from deep underground due to the annihilation processes.       It is much more likely that the direct observations of the axions which will be  inevitably released in the annihilation processes can be directly  observed as recently suggested in  refs \cite{Fischer:2018niu,Liang:2018ecs}. 
 In fact, the observation of these axions with very distinct spectral properties in comparison with conventional galactic axions 
 will be the smoking gun supporting the entire AQN framework, including the proposal on   the primordial Li puzzle  resolution      as advocated in this work.   We finish this work on this positive and optimistic note.

\section*{Acknowledgements} 
 We are tankful to  Edward  Shuryak and John Webb for questions and correspondence. 
 The work of ARZ    was supported in part by the Natural Sciences and Engineering 
Research Council of Canada. 
The work of VVF was supported by  the Australian Research Council.  
 We are thankful to the KITP for organizing the  program ``High Energy Physics at the Sensitivity Frontier"  where this collaboration had started, and which eventually resulted in this work.   This research was supported in part by the National Science Foundation under Grant No. NSF PHY-1748958. 

   \appendix
    \section{Primordial nuclei   fluxes in vicinities of the nuggets \label{flux}}
    The main goal of this Appendix is to argue that the flux of the ions hitting the AQNs surface is sufficiently  large such that a finite portion of all ions from the system will enter the vicinity of  the nuggets during cosmic time $t_0\sim 2\cdot 10^3$s.  This is precisely the assumption as formulated in the text after eq.   (\ref{final1}), and  which is justified a posteriori.
    
    We start with estimation of a number of ions with charge $Z$ entering the vicinity of a single  antinugget per unit time
        \be
        \label{A1}
    \frac{dN_Z}{dt}\sim 4\pi R_{\rm cap}^2 n_{Z} v_Z
    \ee
  where    capture size $R_{\rm cap}$ is  defined by eqs.(\ref{R_cap}) and (\ref{R_cap1}) and $v_Z$ is the ion's velocity in the plasma.  This expression represents a strong underestimation as it does not account for a huge remaining charge 
$Q(R_{\rm cap}) $ at   distance $R_{\rm cap}$ from the nugget. We will correct for this effect at the end of this Appendix.

 We call the  corresponding ions as ``affected" by the presence of AQNs. The number density of the affected ions 
 per unit volume $dV$ per unit time $dt$ can be estimated by multiplying (\ref{A1}) to the density of the antinuggets given by (\ref{n_AQN}), i.e.
   \be
        \label{A2}
    \frac{dN_Z}{dt dV}\sim   n_{\rm AQN}\cdot \frac{dN_Z}{dt}\sim  4\pi R_{\rm cap}^2 n_{\rm AQN}  n_{Z}  v_Z.
    \ee
  We are interested in a relative ratio of the affected ions, rather than in their  absolute values.
  The corresponding ratio is estimated as follows,
   \be
        \label{A3}
   \frac{1}{n_Z} \left(\frac{dN_Z}{ dt dV} \right)\sim   n_{\rm AQN}\cdot \frac{1}{n_Z}  \frac{dN_Z}{dt}\sim  4\pi R_{\rm cap}^2 n_{\rm AQN}   v_Z.~~~~~
    \ee
  Now we want to estimate the total portion  of affected ions by integrating over $\int dt$.
  To simplify the estimates we simply multiply (\ref{A3}) by time  $t_0\sim 2\cdot 10^3$s corresponding to $T\approx 20$~keV because the integral is saturated by the highest possible temperature. The corresponding estimate reads
   \be
        \label{A4}
  \int \frac{dt }{n_Z} \left(\frac{dN_Z}{ dt dV} \right)  \sim  4\pi R_{\rm cap}^2 n_{\rm AQN}   v_Z t_0\sim 0.1,
    \ee
  where for numerical estimates we used parameters for $R_{\rm cap}$ and $n_{\rm AQN}$ defined in the text.
  The estimate (\ref{A4}) implies that at least $10\%$ of all ions from plasma will be affected by the AQNs during the cosmic time $t_0$.
   
   However, as already mentioned, the result (\ref{A4}) should be considered as a strong underestimation of the relevant portion of the affected ions 
   because there is a systematic effect  (yet, not accounted for) due to the presence of a gradient of the residual electric field in the direction to the antinugget as a result of 
   uncompensated charge $Q(R_{\rm cap}) $ at   distance $R_{\rm cap}$ from the nugget as estimated by (\ref{Q_screened1}).   
   
   To account for  the corresponding effect (which obviously enhances the ratio (\ref{A4}))   
   we assume that the residual charge $Q(R_{\rm cap}) $ will be screened on distance $R_{\rm screen}$ determined by the condition\footnote{\label{screening}It is  an underestimation of the  parameter $R_{\rm screen}$   as the exponential tail (when  the  Thomas Fermi approximation breaks down)  is expected to emerge at much larger distances in comparison with simplified formula (\ref{screen}) as mentioned in  Section \ref{estimates}.}   
   \be
   \label{screen}
   Q(R_{\rm cap}) e^{-\frac{(R_{\rm screen}-R_{\rm cap})}{\lambda_D}}\sim 1, ~~ ~~~ \lambda^2_D\simeq \frac{T}{4\pi n_p\alpha}.
   \ee 
  If one uses the numerical parameters for $T, R_{\rm cap}$,  and $n_p$ from the text   
  one arrives to the following estimate for $R_{\rm screen}$ where the residual charge  $Q(R_{\rm cap}) $ is felt by all ions, 
  \be
  \label{R_screen}
  (R_{\rm screen}-R_{\rm cap})\sim 10^{-2} {\rm cm}, 
  \ee
which is obviously  larger than the numerical value for $ R_{\rm cap}$ from (\ref{R_cap2}) which was used in our estimate 
   (\ref{A4}). Taking into  account this effect the estimate (\ref{A4}) is modified and assumes the form 
    \be
        \label{A5}
  \int \frac{dt }{n_Z} \left(\frac{dN_Z}{ dt dV} \right)  \sim  4\pi R_{\rm screen}^2 n_{\rm AQN}   v_Z t_0\sim  1,
    \ee
which implies that the finite portion of all ions of order one  is affected by the AQNs during the cosmic time $t_0$. 
In fact, the  numerical coefficient in (\ref{A5}) is likely to be much larger than one due to our underestimation 
of parameter $R_{\rm screen}$ as mentioned in footnote \ref{screening}. It implies that most of the ions from the system will be entering the vicinities   of the AQNs multiple times  during the cosmic time $t_0$. 

This estimate supports our main conclusion expressed by equations (\ref{final}) and (\ref{final1}) that the finite portion of the Li and Be ions of the entire system will be depleted. This is order of one effect, which is the main claim of this work.


\end{document}